\documentclass[10pt,aps,showpacs,floatfix,twocolumn,amsmath,amssymb,groupedaddress,superscriptaddress]{revtex4}
\usepackage{epsfig}
\usepackage{psfig}
\usepackage{graphicx}
\usepackage{graphics}
\usepackage{xspace}
\usepackage{latexsym}
\usepackage{natbib}
\usepackage{mathrsfs}
\newcommand{\inieq}{\begin{eqnarray}}            
\newcommand{\fineq}{\end{eqnarray}}            
\newcommand{\diff}{{\rm\,d}}                    
\newcommand{\bint}{\mskip .5mu \int \mskip-18mu} 
\newcommand{\be}{\begin{equation}}
\newcommand{\ee}{\end{equation}}
\newcommand{\ba}{\begin{eqnarray}}
\newcommand{\ea}{\end{eqnarray}}

\def\q{\mbox{\boldmath $q$}}

\def\k{\mbox{\boldmath $k$}}

\def\mcv{\mbox{$\mathcal{V}$}}
\begin{document}
\title{Relativistic descriptions of inclusive quasielastic electron scattering: 
application to scaling and superscaling ideas}
\author{Andrea Meucci} 
\affiliation{Dipartimento di Fisica Nucleare e Teorica, 
Universit\`{a} degli Studi di Pavia and \\
Istituto Nazionale di Fisica Nucleare, 
Sezione di Pavia, I-27100 Pavia, Italy}
\author{J.A. Caballero}
\affiliation{Departamento de F\'\i sica At\'omica, Molecular y
Nuclear, Universidad de Sevilla, E-41080 Sevilla, Spain}
\author{C. Giusti}
\affiliation{Dipartimento di Fisica Nucleare e Teorica, 
Universit\`{a} degli Studi di Pavia and \\
Istituto Nazionale di Fisica Nucleare, 
Sezione di Pavia, I-27100 Pavia, Italy}
\author{F.D. Pacati }
\affiliation{Dipartimento di Fisica Nucleare e Teorica, 
Universit\`{a} degli Studi di Pavia and \\
Istituto Nazionale di Fisica Nucleare, 
Sezione di Pavia, I-27100 Pavia, Italy}
\author{J.M. Ud\'{\i}as}
\affiliation{Grupo de F\'{\i}sica Nuclear, Departamento de F\'\i sica At\'omica, Molecular y
Nuclear, Universidad Complutense de Madrid, E-28040 Madrid}

\date{\today}

\begin{abstract}
An analysis of inclusive quasielastic electron scattering is presented using different 
descriptions of the final state interactions within the framework of the relativistic 
impulse approximation. The relativistic Green's function approach is compared with 
calculations based on the use of relativistic purely real mean field potentials in the final state.
Both approaches lead to a redistribution of the strength but conserving the total flux.
Results for the differential cross section at different energies 
are presented. Scaling properties are also analyzed and discussed.
\end{abstract}

\pacs{ 24.10.Jv; 24.10.Cn; 25.30.Fj}

\maketitle


\section{Introduction}
\label{intro}

Electron scattering reactions with nuclei have provided the most detailed and
complete information on nuclear and nucleon structure. Analysis of data for
light-to-heavy nuclei and for different kinematical situations have been
presented in the literature~\cite{book,arc,jourdan,Kel,Walecka}. 
Not only differential cross sections
but also the contribution of the separate response functions have been considered. 
From the theoretical point of view, an important effort has been devoted for the
last 20 years to the description of inclusive and exclusive processes. The high energies
and momenta involved in recent experiments have led to the development of fully relativistic models
describing the scattering 
process~\cite{Chinn89,Ud1,Ud3,Ud4,meucci1,meucci2,meucci3,meucci4,eepv,Jin,Kim,Chiara03}.
Moreover, ingredients beyond the quasielastic (QE)
approach, namely, meson exchange currents, correlations, etc., have been also considered
following different 
approaches~\cite{Orden,Gross,BCDM04,Amaro:2001xz,Amaro:2002mj,Amaro:2003yd,DePace,ryck3,bright1,bright2,co,mcder,mecpvnr,mecpvnr1,mecpv}.
 
Within the QE domain, which is the region
considered in this work, the treatment of final state
interactions (FSI) between the ejected nucleon and the residual nucleus 
has been proved to be essential in order
to compare with data. This has been well established in the case of exclusive
$(e,e'N)$ reactions where an ejected nucleon is detected in coincidence with the scattered
electron. Analyses based on the use of phenomenological complex potentials in the final
channel have provided results in accordance with data~\cite{book,Kel}. In particular, the use of
relativistic complex optical potentials within a fully relativistic description of
the reaction mechanism has led to theoretical cross sections in excellent accordance with
data~\cite{Ud1,Ud3,Jin,Kim,meucci1,meucci2,meucci4}. 
Moreover, comparison with separate responses and asymmetries has also proved the
capability of the relativistic approach to successfully describe fine details of
data behavior~\cite{Ud4,Udi-PRL,Cris04,meucci1,meucci4}.

In the analysis of inclusive reactions, contrary to exclusive ones, all inelastic channels
in the final state should be retained. This means that the flux is conserved, and
consequently, the distorted wave impulse approximation (DWIA) based on the use of a complex potential
should be dismissed. However, final state interaction continues to be a main 
ingredient in the inclusive process, and its appropriate description is required in order to describe data.
Within the framework of fully relativistic models different approaches have been presented
in the literature.
On the one hand, a description based on the relativistic distorted
wave impulse approximation (RDWIA) has been pursued, but using purely real potentials in the
final channel. This is consistent with flux conservation. Concerning the specific potentials, 
the final
nucleon state has been evaluated with the real part of the relativistic energy-dependent optical
potential, denoted as rROP, or with the same relativistic mean
field potential considered in describing the initial nucleon state, 
denoted as RMF (see~\cite{Jin,Kim,cab1,Chiara03}).
A second approach of FSI makes use of the relativistic Green's function technique, where
the components of the nuclear response are written in terms of the single particle optical
model Green's function. This method, denoted as GF, allows one to perform calculations treating FSI consistently
in the inclusive and exclusive channels. The same relativistic (complex) optical potential
is considered in both cases, but flux is conserved in the inclusive process. Moreover,
redistribution of strength among different channels is due to the real and also 
significantly to the imaginary part of the potential (see~\cite{eepv,ee,cc,eenr,eeann}
for details).

The exhaustive analysis of the $(e,e')$ world data has demonstrated the quality of
scaling arguments at high momentum transfer for excitation energies below the QE 
peak~\cite{mai1,don1,don2,Day90}.
These data, and particularly those coming from the separate longitudinal contribution, 
have been shown to respect scaling of first kind (independence of the
transferred momentum $q$)
and scaling of second kind (no dependence on the nuclear system).
The analysis of the longitudinal response, very mildly affected by meson exchange currents and
nuclear correlations, has permitted the extraction of a phenomenological QE superscaling
function given for transferred energies below and above
the QE peak. These regions correspond to negative and positive values of the so-called
superscaling variable, respectively~\cite{mai1,don1,don2,cab2,cab1}. 
Scaling analysis for electron scattering, which was extended
into the $\Delta$ region~\cite{BCDM04,amaro05,ivanov08}, 
has been also exploited to predict inclusive 
charged-current (CC) neutrino-nucleus 
cross sections~\cite{cab1,cab2,amaro05,isospin07,neutrino2,antonov06,Martini:2007jw} 
and neutral-current (NC) processes~\cite{NC,nccris,antonov07}.

The QE phenomenological scaling function presents a
significant asymmetry with a tail extended to high values of the transferred energy $\omega$
(positive values of the scaling variable $\psi'(q,\omega)$). 
This imposes strong restrictions to all theoretical models describing QE $(e,e')$
processes, namely, they should be able to fulfill scaling properties 
and in addition reproduce the specific shape of the scaling function. 
Asymmetry in the
scaling function is largely absent in non-relativistic models based on a mean field 
approach~\cite{neutrino2}. The same comment applies to 
the relativistic plane wave limit and to results
based on a DWIA using real relativistic energy-dependent optical 
potentials~\cite{cab1,cab2}. In contrast, asymmetry comes out within the relativistic impulse
approximation, but with FSI described using strong relativistic
mean field  potentials. An asymmetrical scaling function has been also shown to occur
within the framework of a semi-relativistic model, based on FSI given through the Dirac
equation-based (DEB) potential derived from the RMF (see~\cite{amaro07} for details). 

Previous studies clearly illustrate the central role played by FSI in providing 
theoretical results in accordance with data. Hence, in this work we present a
careful analysis of $(e,e')$ processes comparing two basic descriptions of FSI:
the RMF approach and the relativistic GF technique. Both models have
their own merits. In the RMF the flux lost into inelasticities is recovered by using
the same, purely real, energy-independent potential seen by the bound nucleons and thus no
information from scattering reactions is explicitly incorporated in the model.
In this way it is consistent with dispersion relations~\cite{hori}.
The RMF leads to an asymmetrical scaling
function which is supported on overall by data behavior. However, it 
breaks down scaling significantly as the transferred momentum $q$ increases,
particularly in the region above the QE peak. The analysis of experimental $(e,e')$ data
indeed leaves room for first-kind scaling breaking in this region, due partly to
$\Delta$ production and other contributions beyond the impulse approximation, namely,
meson exchange currents and their impact in the $2p-2h$ sector~\cite{DePace}.
The GF approach provides a consistent and unified treatment of inclusive 
and exclusive electron-nucleus scattering processes. 
It also fullfills dispersion relations and 
recovers the flux lost into inelasticities by means of a formal summation of unobserved channels performed
via the optical potential. Most often, the optical potential is taken from phenomenology, and thus,
in this approach, information on the observed nucleon-nucleus scattering is 
incorporated into the model.

The differences observed between the predictions of the two approaches and 
their behaviour with regard to scaling arguments will be helpful in settling down 
an appropriate description of FSI for inclusive $(e,e')$ processes at different kinematics. 
Further, this would be useful for understanding nuclear effects, 
especially FSI, in neutrino-nucleus 
cross-sections \cite{cab1,cab2,amaro05,cc,Cris06,Lava,Meucci:2006ir,Meucci:2008zz,
Butkevich:2007gm,Butkevich:2009cp,Benhar:2006nr,Leitner:2006ww,Buss:2007ar,Leitner:2008ue}, 
which are of paramount interest for the ongoing and future neutrino oscillation 
experiments \cite{nuint,k2k,miniboone,sciboone,MicroBooNE,t2k,minos,minerva,nova}. 

Therefore, in this work we present a systematic study of both models,
comparing their predictions and analyzing their scaling properties,
with special emphasis on the specific shape of the scaling function.

The paper is organized as follows. In Sec. II, we present the basic formalism involved
in inclusive quasielastic electron scattering with a discussion of the different
models considered in the description of FSI, namely, the relativistic mean field
and the relativistic Green's function approach. We also provide a subsection which contains
the basic facts related to scaling arguments and introduce superscaling functions.
In Sec. III, we show and discuss our results. First, we focus on the analysis of
differential cross sections and later on we analyze in detail results for the 
scaling function. Finally, in Sec. IV we summarize the main results and present
our conclusions.


\section{Inclusive quasielastic electron scattering}
\label{sec.for}

An electron with four-momentum $K^{\mu} = (\varepsilon,\k)$ is scattered through
an angle $\vartheta_e$ to four-momentum 
$K^{\prime\mu} = (\varepsilon^{\prime},\k^{\prime})$. The four-momentum transfer 
is $Q^{\mu} = K^{\mu} -  K^{\prime\mu} = (\omega,\q)$.
In the one-photon exchange approximation the inclusive cross section for the
quasielastic $(e,e^{\prime})$ scattering on a nucleus is given by~\cite{book}
\inieq
\left(\frac{\diff \sigma}{\diff \varepsilon^{\prime} \diff \Omega^{\prime}}\right) =
\sigma_{{M}} 
\left[ V^{}_{{L}}R_{{L}} +  V^{}_{{T}}R_{{T}}\right] 
\ , \label{eq.cspc}
\fineq
where $\Omega^{\prime} $ is the scattered electron solid angle and 
$\sigma_{{M}}$ is the Mott cross section~\cite{book}. The 
coefficients $V$ come from the lepton tensor components and are defined as
\begin{eqnarray}
V^{}_{{L}} = \left(\frac {|Q^2|} {|\q|^2}\right)^2 \ , \ 
V^{}_{{T}}=\tan^2\frac {\vartheta_e}{2} - \frac{|Q^2|} {2|\q|^2} \  ,\ 
\label{eq.lepton}
\end{eqnarray}
where $|Q^2| = |\q|^2 - \omega^2$.
 
All nuclear
structure information is contained in the longitudinal and transverse response
functions, $R_{{L}}$ and $R_{{T}}$, expressed by
\inieq
R_{{L}}(q,\omega) &=& W_{\textrm{}}^{00}(q,\omega) \ , \nonumber \\
R_{{T}}(q,\omega) &=& W_{\textrm{}}^{11}(q,\omega) + 
      W_{\textrm{}}^{22}(q,\omega) \ ,
\label{eq.response}
\fineq
in terms of the diagonal components of the hadron tensor
that is given by bilinear products of the transition matrix
elements of the nuclear electromagnetic many-body current operator $\hat{J}^{\mu}$ between
the initial state $\mid\Psi_0\rangle$ of the nucleus, of energy $E_0$, and the 
final states $\mid \Psi_{\textrm {f}} \rangle$, of energy $E_{\textrm {f}}$, 
both eigenstates of the $(A+1)$-body Hamiltonian $H$, as 
\begin{eqnarray}
 W^{\mu\nu}(q,\omega) &=& \overline{\sum}_i
 \bint\sum_{ {f}}  \langle 
\Psi_{ {f}}\mid \hat{J}^{\mu}(\q) \mid \Psi_0\rangle \nonumber \\ &\times&
\langle 
\Psi_0\mid \hat{J}^{\nu\dagger}(\q) \mid \Psi_{ {f}}\rangle 
\ \delta (E_0 +\omega - E_{ {f}}),
\label{eq.ha1}
\end{eqnarray}
involving an average over the initial states and a sum over the undetected final 
states. The sum runs over the scattering states corresponding to all of the 
allowed asymptotic configurations and includes possible discrete states.  

In the QE region the nuclear response is dominated by one-nucleon
knockout processes, where the scattering occurs with only one nucleon which is
subsequently emitted. The remaining nucleons of the target behave as simple spectators.
Therefore, QE electron scattering is adequately described in 
the Relativistic Impulse Approximation (RIA) by the sum of incoherent
processes involving only one nucleon scattering.

In the RIA the nuclear current operator is assumed to be the sum 
of single-nucleon currents $\hat{j}^{\mu}$, for which different relativistic free nucleon
expressions~\cite{deF} can be used. In the present calculations we use the
option denoted as CC2, i.e., 
\begin{eqnarray}
  \hat{j}^{\mu} &=&  F_1(Q^2) \gamma ^{\mu} + 
             i\frac {\kappa}{2M} F_2(Q^2)\sigma^{\mu\nu}Q_{\nu}
	     \ ,
	     \label{eq.cc}
\end{eqnarray}
where $\kappa$ is the anomalous part of 
the magnetic moment and
$\sigma^{\mu\nu}=\left(i/2\right)\left[\gamma^{\mu},\gamma^{\nu}\right]$.
$F_1$ and $F_2$ are the Dirac and Pauli 
nucleon form factors~\cite{gal71}.


Within the RIA framework and under the assumption of a shell-model description
for nuclear structure,
the components of the hadron tensor are obtained from the sum, over all the 
single-particle (s.p.) shell-model states, of the squared absolute value of the transition 
matrix elements of the single-nucleon current
\inieq
\langle\chi_{{E}}^{(-)}(E)\mid  \hat{j}^{\mu}(\q)\mid \varphi_n \rangle  \ ,
	     \label{eq.dko}
\fineq
where $\chi_{{E}}^{(-)}(E)$ is the scattering state of the emitted nucleon and 
the overlap $\varphi_n$ between the ground state of the target 
$\mid\Psi_0\rangle$ and the final state $ \mid n\rangle$ of the residual 
nucleus is a s.p. shell-model state.
 
In the calculations presented in this work the bound nucleon states $\varphi_n$ are taken as 
self-consistent Dirac-Hartree solutions derived within a RMF
approach using a Lagrangian containing $\sigma$, $\omega$ and $\rho$ 
mesons~\cite{boundwf,Serot,adfx,lala,sha}. 

Different prescriptions are used to calculate the relativistic s.p. scattering 
wave functions. 
In the simplest approach the plane-wave limit is considered, i.e., the
Relativistic Plane-Wave Impulse Approximation (RPWIA), where FSI 
between the outgoing nucleon and the residual nucleus are neglected.
In the approaches based on the relativistic distorted-wave impulse 
approximation  FSI effects are accounted for by solving the Dirac 
equation with relativistic optical potentials. The use of relativistic 
energy-dependent complex optical potentials fitted to elastic proton-nucleus
scattering data has been very successfull in describing exclusive $(e,e^{\prime}p)$  
data~\cite{book,Ud1,Ud3,Ud4,Kel,meucci1,meucci2,meucci3,meucci4}. 
In the exclusive scattering the imaginary part of the optical potential produces an
absorption that reduces the calculated cross section and accounts for the flux
lost in the considered elastic channel towards other channels. This 
approach is appropriate for the exclusive process where only one channel 
is considered, but it would be inconsistent for the inclusive scattering, where 
all the inelastic channels should be retained and the total flux, although
redistributed among all possible channels due to FSI,
must be conserved. As a result, in the RDWIA (with complex potentials) the flux is not conserved 
and the inclusive $(e,e^{\prime})$ cross section is 
underestimated~\cite{Chiara03,Jin,Kim,ee}. 
A simple way of estimating the inclusive response within the RIA, avoiding 
spurious flux absorption, is to use purely real potentials. In a first approach, the imaginary 
part of the phenomenological relativistic energy-dependent optical  
potentials~\cite{chc} is set to zero, thus retaining in the calculations only 
the real part. In a second approach, the scattering states are described by 
using the same RMF theory potentials considered in the description of 
the initial bound state $\varphi_n$. We refer to these two different 
FSI descriptions as real Relativistic Optical Potential (rROP) and 
Relativistic Mean Field (RMF), respectively.

We note that rROP conserves the flux and thus it is inconsistent with
the exclusive process, where a complex optical potential must be used. Moreover,
the use of a real optical potential is unsatisfactory from a theoretical point
of view, since it is an energy-dependent potential, reflecting the different contribution of 
open inelastic channels for each energy. Dispersion relation then dictates that the 
potential should have a nonzero imaginary term~\cite{hori}.
On the other hand,
the RMF model is based on the use of the same strong energy-independent real 
potential for both bound and scattering states, so that it fulfills the
dispersion relation~\cite{hori} and, further, it maintains the continuity equation.

Green's function techniques are exploited to derive the inclusive response 
in a different model where the flux is conserved and the use of a complex optical 
potential allows one to treat FSI consistently in the inclusive and in the 
exclusive reactions. Detailed discussions of this Green's function (GF) approach
can be found in~\cite{ee,cc,eepv,eenr,eeann}. Here we recall only the essential 
features of the model. 

In the GF approach the components of the nuclear response in Eq.~(\ref{eq.ha1}) are 
written in terms of the s.p. optical model Green's function. This is the 
result of suitable approximations, such as the one-body current assumption 
and subtler approximations related to the IA. 
The explicit calculation of the s.p. Green's function can be avoided by its spectral 
representation, which is based on a biorthogonal expansion in terms of a non 
Hermitian optical potential and of its Hermitian conjugate. 
The nuclear response of Eq.~(\ref{eq.ha1}) is then obtained in the form~\cite{ee}
\inieq
W^{\mu\mu}(q,\omega) = \sum_n \Bigg[ \textrm{Re} T_n^{\mu\mu}
(E_{\textrm{f}}-\varepsilon_n, E_{\textrm{f}}-\varepsilon_n)  \nonumber
\\
- \frac{1}{\pi} \mathcal{P}  \int_M^{\infty} \diff \mathcal{E} 
\frac{1}{E_{\textrm{f}}-\varepsilon_n-\mathcal{E}} 
\textrm{Im} T_n^{\mu\mu}
(\mathcal{E},E_{\textrm{f}}-\varepsilon_n) \Bigg] \ , \label{eq.finale}
\fineq
where $n$ denotes the eigenstate of the residual nucleus related to the 
discrete eigenvalue $\varepsilon_n$ and 
\inieq
T_n^{\mu\mu}(\mathcal{E} ,E) &=& \langle \varphi_n
\mid \hat{j}^{\mu\dagger}(\q) \sqrt{1-\mcv'(E)}
\mid\tilde{\chi}_{\mathcal{E}}^{(-)}(E)\rangle \nonumber \\
&\times&  \langle\chi_{\mathcal{E}}^{(-)}(E)\mid  \sqrt{1-\mcv'(E)} \hat{j}^{\mu}
(\q)\mid \varphi_n \rangle  \ . \label{eq.tprac}
\fineq
The factor $\sqrt{1-\mcv'(E)}$, where $\mcv'(E)$ is the energy derivative of 
the optical potential, accounts for interference effects between different 
channels and justifies the replacement in the calculations of the Feshbach 
optical potential $\mcv$ of the GF model, for which neither microscopic nor
empirical calculations are available, by the local phenomenological optical 
potential~\cite{ee,eenr}. 
Disregarding the square root correction, the second matrix element in 
Eq.~(\ref{eq.tprac}) is the transition amplitude of single-nucleon knockout of 
Eq.~(\ref{eq.dko}), where the imaginary part of the optical potential
accounts for the flux lost in the channel $n$  towards the channels different 
from $n$. In the inclusive response this loss is compensated by a corresponding 
gain of flux due to the flux lost, towards the channel $n$, by the other final 
states asymptotically originated by the channels different from $n$. 
This compensation is performed by the first matrix element in the right hand 
side of Eq.~(\ref{eq.tprac}), that is similar to the matrix element of 
Eq.~(\ref{eq.dko}) but involves the eigenfunction 
$\tilde{\chi}_{\mathcal{E}}^{(-)}(E)$ of the Hermitian conjugate optical
potential, where the imaginary part has an opposite sign and has the 
effect of increasing the strength. In the GF approach the imaginary part of the optical
potential redistributes the flux lost in a channel in the other channels, and 
in the sum over $n$ the total flux is conserved.  

The hadron tensor in Eq.~(\ref{eq.finale}) is the sum of two terms. The 
calculation of the second term requires integration over all the eigenfunctions of the
continuum spectrum of the optical potential. 
If the imaginary part of the
optical potential is neglected, the second term in Eq.~(\ref{eq.finale}) vanishes
and, but for the square root factor (whose contribution is however generally 
small in the calculations), the first term gives the rROP approach.

\subsection{Scaling at the quasielastic peak}
\label{sec.sca}

Scaling ideas applied to inclusive QE electron-nucleus scattering
reactions have been shown to work properly to high accuracy~\cite{mai1,don1,don2}.
In fact, an exhaustive analysis of QE $(e,e')$ world data has demonstrated that scaling
property is well respected at high momentum transfer for excitation energies falling
below the QE peak. These data, when plotted against a properly chosen
variable (the scaling variable), show
a mild dependence on the momentum transfer (reasonable scaling of first kind) 
and almost no dependence on the nuclear target (excellent scaling of second kind).
Simultaneous fulfillment of both conditions leads to superscaling. The analysis of
the separated longitudinal contribution to $(e,e')$ data has led to introduce a
phenomenological scaling function which contains the relevant information about
the nuclear dynamics explored by the probe. Scaling and superscaling are general
phenomena exhibited by nature which any ``reliable'' model providing a description of 
the scattering process should be able to reproduce. Not only superscaling
behavior should be described,
but also the specific magnitude and shape of the ``universal'' phenomenological 
superscaling function needs to be reproduced.    

The usual procedure to get the scaling function consists of dividing the
inclusive differential cross section (\ref{eq.cspc}) by the
appropriate single-nucleon $eN$ elastic cross section weighted by the 
corresponding proton and neutron numbers~\cite{mai1,don1,don2,Barbaro:1998gu} 
involved in the process,
\be
f(\psi',q)\equiv k_F\frac{\left[\displaystyle\frac{d\sigma}{d\varepsilon'd\Omega'}\right]}
{\sigma_M\left[V_LG_L(q,\omega)+V_TG_T(q,\omega)\right]} \, .
\label{fscaling}
\ee
Analogously, the analysis of the separated longitudinal ($L$) and transverse ($T$)
contributions leads to scaling functions,
\be
f_{L(T)}(\psi',q)\equiv k_F\frac{R_{L(T)}(q,\omega)}{G_{L(T)}(q,\omega)} 
\label{fltscaling} \, .
\ee
The term $\psi'(q,\omega)$ is the dimensionless scaling variable extracted
from the relativistic Fermi gas (RFG) analysis that incorporates the typical momentum scale for the selected
nucleus~\cite{mai1,cab1}.
The fully relativistic expressions for $G_{L(T)}$ involve the proton and neutron 
form factors $G_{E(M)}^{pn}$, weighted by the proton and neutron numbers, and
an additional dependence on the nuclear scale given through the Fermi momentum $k_F$
(explicit expressions are given by Eqs.~(16-19) in~\cite{mai1}). 

Whereas $L$ data are compatible with superscaling behavior, permitting the extraction of the phenomenological
function $f_L^{exp}(\psi')$, scaling is known to be violated in the $T$ channel
at energies above the QE peak by effects beyond the impulse 
approximation~\cite{Alvarez-Ruso:2003gj,BCDM04,Amaro:2001xz,Amaro:2002mj,Amaro:2003yd,DePace}. 
It is important to point out that, although many models based on the impulse approximation
exhibit superscaling, even perfectly as the RFG, only a few
of them are capable to accurately reproduce the asymmetric shape of $f_L^{exp}$ with a significant
tail extended to high transferred energies (large positive values of the scaling variable $\psi'$).
One of these is based on the RIA with final state interactions described by 
means of strong relativistic mean field potentials.
On the contrary, calculations based on the plane wave limit and/or the use of real
relativistic optical potentials  in the final state, although satisfying superscaling
properties, lead to ``symmetrical-shape'' scaling functions which are not in accordance 
with data analysis~\cite{cab2,cab1,isospin07}.

In this work we extend the analysis of scaling to the
relativistic GF approach, whose predictions will be compared with previous
results obtained within the RMF, RPWIA, and rROP frameworks. This study will show to
which degree superscaling is fulfilled by the GF calculations, and, moreover, how the
GF scaling function compares with the experiment. As already mentioned, the GF treats
FSI consistently in the inclusive and exclusive reactions, hence its application to scaling
and superscaling ideas, which emerge as an essential outcome of nature, results mandatory.
This allows us to get a clear insight concerning the capability of the
relativistic GF approach
applied to QE inclusive $(e,e')$ reactions, as well as the uncertainties ascribed to the 
various ingredients of the model, particularly the specific energy-dependent terms in the
complex optical potentials involved.

\begin{figure}[ht]
\begin{center}
\includegraphics[height=9cm, width=9cm]{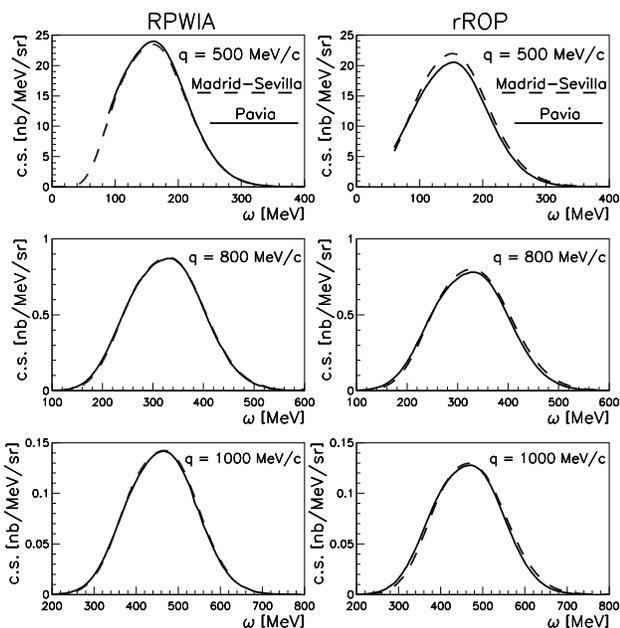} 
\vskip -0.6cm
\caption { Differential cross section of the $^{12}$C$(e,e^{\prime})$ reaction 
for an incident electron energy $\varepsilon = 1$ GeV and three values of the 
momentum transfer, i.e., $q = $ 500, 800, and 1000 MeV$/c$, calculated by the 
Pavia (solid) and the Madrid-Sevilla (dashed) groups with RPWIA (left) and 
rROP (right). 
}
\label{fig1}
\end{center}
\end{figure}
\begin{figure}[ht]
\begin{center}
\includegraphics[height=9cm, width=9cm]{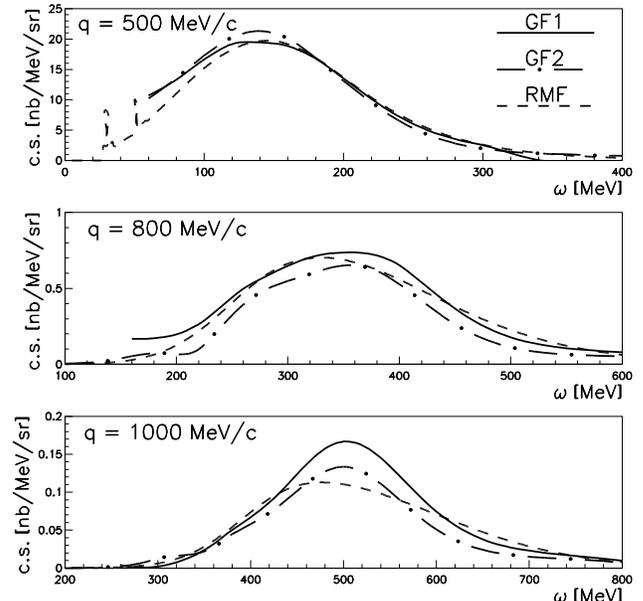} 
\vskip -0.6cm
\caption { Differential cross section of the $^{12}$C$(e,e^{\prime})$ reaction 
in the same kinematics as in  Fig. \ref{fig1}.
The solid and long dot-dashed lines are the GF results calculated with the 
two different optical potentials EDAD1 and EDAD2. The dashed lines are the 
results of the RMF model.
}
\label{fig2}
\end{center}
\end{figure}
\begin{figure}[ht]
\begin{center}
\includegraphics[height=9cm, width=9cm]{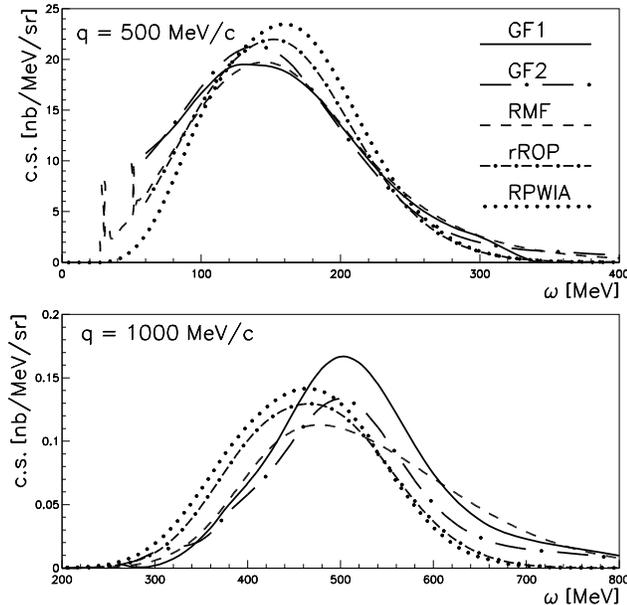} 
\vskip -0.6cm
\caption { Differential cross section of the $^{12}$C$(e,e^{\prime})$ reaction 
for $\varepsilon = 1$ GeV and $q = $ 500 and 1000 MeV$/c$. The solid, 
long dot-dashed, and dashed lines are the same as  in Fig. \ref{fig1}.
The dot-dashed and dotted lines are the rROP and RPWIA results, respectively.
}
\label{fig3}
\end{center}
\end{figure}
\begin{figure}[ht]
\begin{center}
\includegraphics[height=9cm, width=9cm]{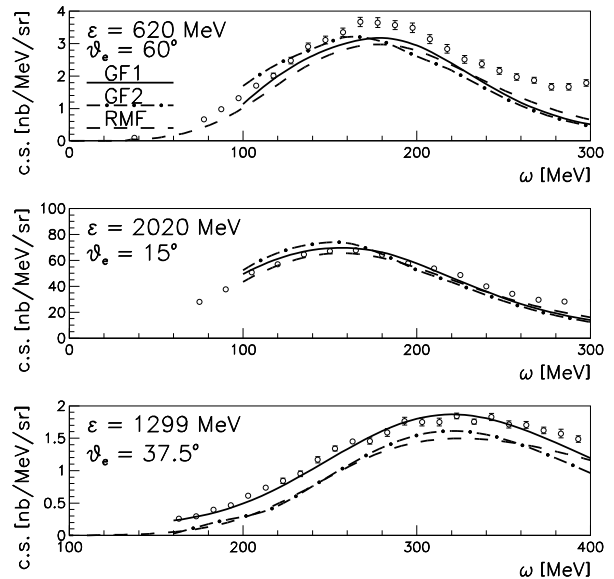} 
\vskip -0.6cm
\caption { Differential cross section of the $^{12}$C$(e,e^{\prime})$ reaction 
for different beam energies and electron scattering angles. Line convention as in 
Fig. \ref{fig2}, experimental data from~\cite{620,day,1299}. 
}
\label{fig4}
\end{center}
\end{figure}
\begin{figure}[ht]
\begin{center}
\includegraphics[height=9cm, width=9cm]{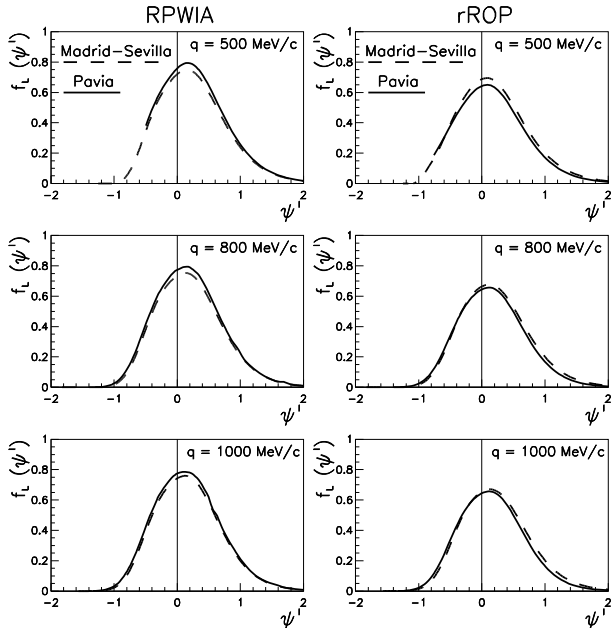} 
\vskip -0.6cm
\caption { Longitudinal contribution to the scaling function for three 
values of the momentum transfer, i.e., $q = $ 500, 800, and 1000 MeV$/c$, 
obtained by the 
Pavia (solid) and the Madrid-Sevilla (dashed) groups with RPWIA (left) and 
rROP (right). 
}
\label{fig5}
\end{center}
\end{figure}
\begin{figure}[ht]
\begin{center}
\includegraphics[height=9cm, width=9cm]{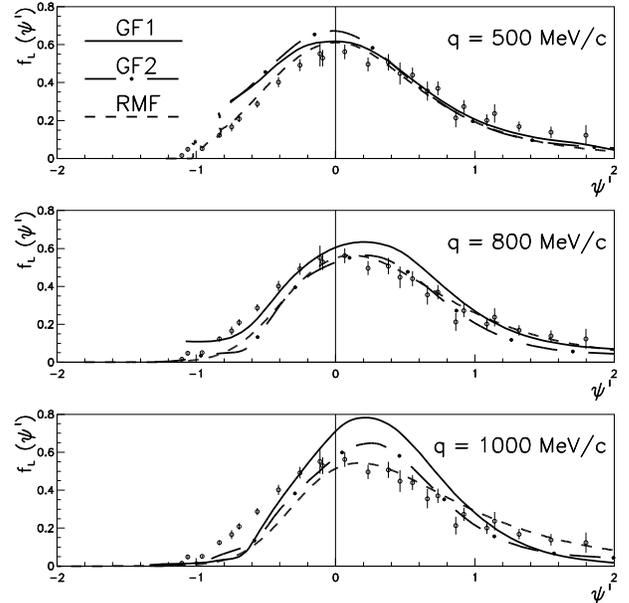} 
\vskip -0.6cm
\caption {  Longitudinal contribution to the scaling function for  $q = $ 500, 
800, and 1000 MeV$/c$  with the GF1 (solid), GF2 
(long dot-dashed), and RMF (dashed) models compared with the averaged 
experimental scaling function.
}
\label{fig6}
\end{center}
\end{figure}
\begin{figure}[ht]
\begin{center}
\includegraphics[height=9cm, width=9cm]{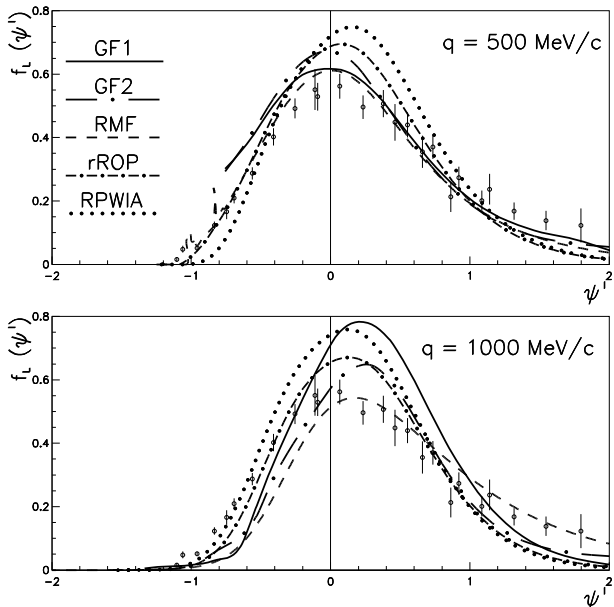} 
\vskip -0.6cm
\caption { Longitudinal contributions to the scaling function for $q = 500$ 
and $1000$ MeV$/c$ compared with the averaged experimental scaling function. 
Line convention as in Fig. \ref{fig3}. 
}
\label{fig7}
\end{center}
\end{figure}
\begin{figure}[ht]
\begin{center}
\includegraphics[height=9cm, width=9cm]{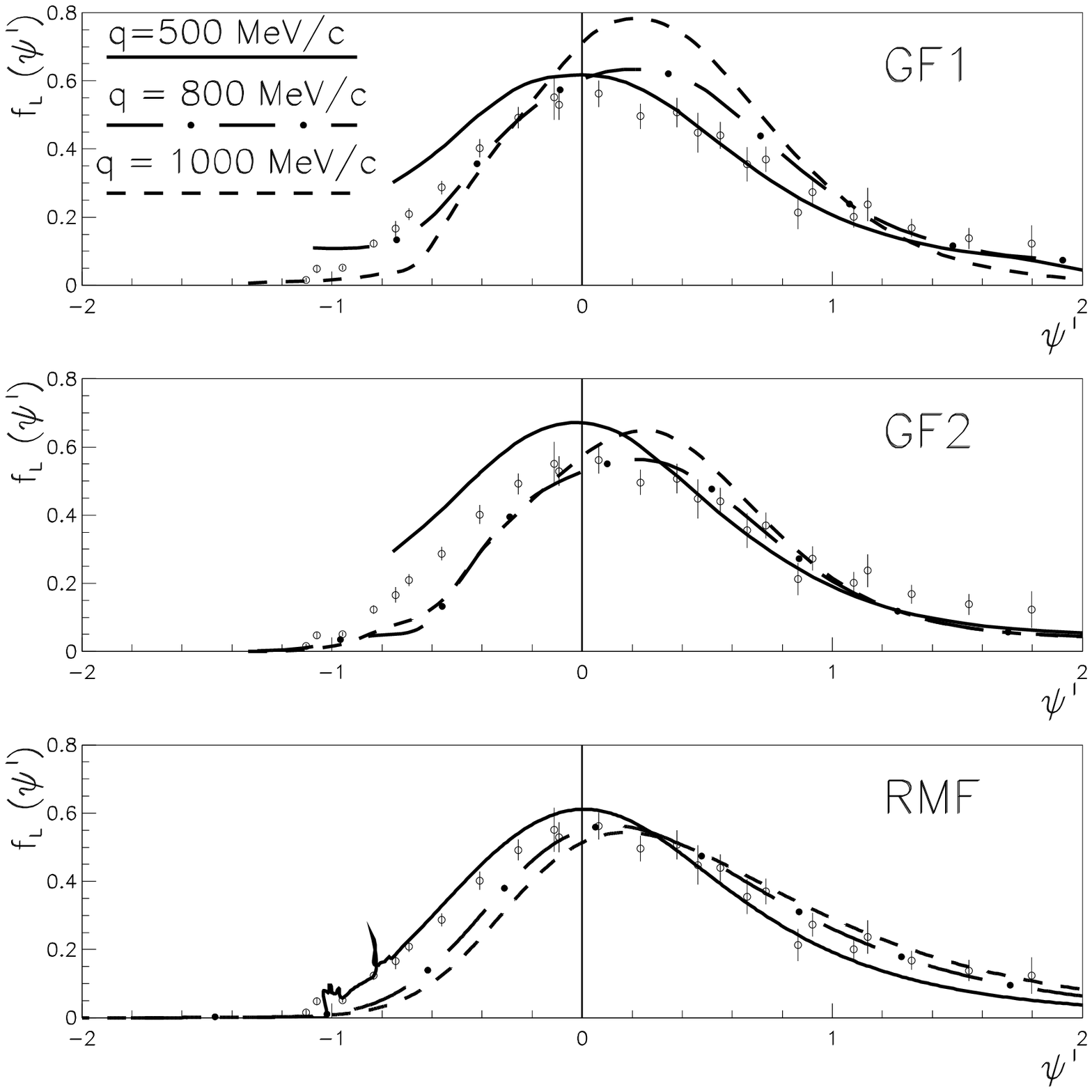} 
\vskip -0.6cm
\caption { Analysis of first-kind scaling, 
$f_{{L}}(\psi^{\prime})$ for  $q = 500$ (solid), $q = 800$ (dot-dashed), 
and $1000$ MeV$/c$ (dashed) with the GF1 (top panel), GF2 (middle 
panel), and RMF (bottom panel) models using the same results already displayed 
in Fig. \ref{fig6}.
}
\label{fig9}
\end{center}
\end{figure}

\section{Results and discussion}
\label{results}

In this section the numerical results of the different relativistic models developed
by the Pavia and the Madrid-Sevilla groups to describe FSI in the inclusive
quasielastic electron-nucleus scattering are considered. As a first step
results obtained by the two groups in the RPWIA and rROP approaches are compared
in order to check the consistency of the numerical programs when
calculations are carried out under the same conditions. Then the results 
corresponding to the RMF model developed by the Madrid-Sevilla group and the relativistic 
GF model developed by the Pavia group are compared. This comparison is performed for the 
$^{12}$C$(e,e^{\prime})$ cross section and scaling function calculated with various
models under different kinematics.

\subsection{Differential cross section}
  
The $^{12}$C $(e,e^{\prime})$ cross section has been calculated for a fixed 
value of the incident electron energy $\varepsilon = 1$ GeV and three 
values of the momentum transfer, i.e. $q = $ 500, 800 and
1000 MeV$/c$. The relativistic initial states are taken as Dirac-Hartree 
solutions of a relativistic Lagrangian written in the context of a relativistic 
mean field theory with the NLSH parameterization~\cite{adfx,lala,sha}. 

The cross sections calculated in RPWIA and in rROP by the Pavia and the 
Madrid-Sevilla groups are compared in Fig.~\ref{fig1}. 
The calculations are performed using the same ingredients for the relativistic
wave-functions and the current operator. Almost identical results are obtained 
in RPWIA. In rROP the two results are very similar, up to a few percent.

The comparison in Fig.~\ref{fig1} is a first important and necessary 
benchmark of the two independent computer programs, which 
allows us to estimate the numerical uncertainties and 
gives enough confidence on the reliability of both calculations. Further discrepancies
between the results of the two groups can now be ascribed to differences of 
the models but not to inconsistencies in the calculations.

The cross sections evaluated with the RMF and GF models for the same 
kinematics as in Fig.~\ref{fig1} are presented in Fig.~\ref{fig2}.
In the case of the GF approach two different parameterizations for the relativistic 
optical potential have been used: the energy-dependent and A-dependent EDAD1 
and EDAD2 complex phenomenological potentials of~\cite{chc}, which are fitted 
to proton elastic scattering data on several nuclei in an energy range up to 
1040 MeV. In the figures the results obtained with EDAD1 and EDAD2 are denoted 
as GF1 and GF2, respectively. 
The differences between the results of the RMF and GF models increase with the
momentum transfer. Also discrepancies between the GF1 and GF2 cross 
sections depend on the momentum transfer.  
At $q$ = 500 MeV$/c$ the three results in Fig.~\ref{fig2} are similar, both in
magnitude and shape. At larger $q$, $q$ = 800 MeV$/c$, moderate differences are 
found, whereas the discrepancy between the three approaches gets larger
at $q = 1000$ MeV$/c$.
The shape of the RMF cross section shows an asymmetry, with a long tail 
extending towards higher values of $\omega$, that is essentially due to the 
strong energy-independent scalar and vector potentials present in the RMF approach.
On the contrary, in the case of GF1 and GF2, the asymmetry towards higher $\omega$ is 
less significant but still visible. The GF1 and GF2 cross sections show a
similar shape but with a significant difference in the magnitude. 
At $q$ = 1000 MeV$/c$ both of them 
are higher than the RMF cross section in the region where the maximum occurs.
However, note that a stronger enhancement is obtained with GF1, which at the peak 
overshoots the RMF cross section up to $40\%$.
Overall, taking into account the different ingredients that these estimates of the inclusive response 
contain, the similarity of the predictions for the inclusive cross-section is remarkable, particularly at
the two lower values of $q$. The larger differences seen for the largest $q$ value, 
not only between the RMF and GF models, but also between the two GF results, is 
simply an indication of the difference in the ingredients of these calculations.

Indeed, the different behaviour presented by the RMF and GF results as a function of $q$ and
$\omega$ is directly linked to the specific structure of the relativistic potentials
involved in the RMF and GF models. Whereas the RMF is based on the 
use of a strong energy-independent real potential, the GF approach makes use of a complex 
energy-dependent optical potential. In GF calculations the behavior of the 
optical potential changes with the momentum and energy transferred
in the process, and higher values 
of $q$ and $\omega$ correspond to higher energies for the optical potential. 
The results obtained with GF1 and GF2 are consistent with the general 
behaviour of the phenomenological relativistic optical potentials 
and are basically due to their imaginary part.
To make this clear, we present again in Fig.~\ref{fig3} the GF and RMF cross sections 
for $q = $ 500 and 1000 MeV$/c$ compared directly with the corresponding results obtained 
within the RPWIA and rROP models. It is known that the real terms of the relativistic
optical potentials are very similar for the different parameterizations. This explains why
the cross section evaluated within the rROP approach does not show sensitivity to 
the particular parameterization considered for the ROP. On the other hand,
the energy-dependent scalar and vector components of the real part of the ROP get smaller
with increasing energies. Thus the rROP result approaches the RPWIA one for large values of $\omega$.
In contrast, the imaginary (scalar and vector) part presents its maximum strength around 500 MeV 
being also sensitive to the particular ROP parameterizations. This explains the differences observed
between the rROP and the two GF results as a function of $\omega$ and $q$. 
This significant discrepancy between GF and rROP cross sections in 
Fig.~\ref{fig3} seems to contradict previous results shown in~\cite{cc}. However,
kinematical conditions reported in~\cite{cc} corresponded to lower
values of the momentum transfer, where there is no reason {\it a priori} to expect 
the rROP and GF predictions to be closer or further apart.

Of particular relevance is the difference between the
GF1 and GF2 results. These are obtained with optical potentials that reproduce 
the elastic proton-nucleus phenomenology
to a similar degree~\cite{chc}. However, one must keep in mind that elastic observables do not completely constrain
optical potentials, and indeed it has been often seen how the predictions of the EDAD1 and EDAD2 potentials
for nonelastic observables, such as for instance $(e,e'p)$ or electron
trasparencies~\cite{Cris06,Lava}, 
differ significantly. The differences between GF1 and GF2 are mostly due  
to their different imaginary part, that includes the overall effect of the
inelastic channels and is not univocally determined by the 
elastic phenomenology.  The most convenient choice of the phenomelogical 
optical potential to be employed within the GF approach should thus be made after a comparison to inclusive data.

In Fig.~\ref{fig4} the GF1, GF2, and RMF results are compared with the
experimental cross sections for three different kinematics~\cite{620,day,1299}. 
A recent review of the experimental situation as well as different theoretical 
approaches can be found in~\cite{arc}. Results in Fig.~\ref{fig4} show that the
three models lead to similar cross sections. The main differences are presented 
for higher values of the momentum transfer, about 800 MeV/$c$ (bottom panel),
where the GF1 cross section (solid line) is larger than the GF2 (dot-dashed) 
and RMF (dashed) ones. The experimental cross section is slightly underpredicted 
in the top panel and well described in the middle panel by all calculations. 
Finally, results in the bottom panel show a fair agreement with data in the case of
GF1, whereas GF2 and RMF underpredict the experiment. Summarizing, the comparison with
data, although satisfactory on general grounds, gives only an indication and cannot be conclusive 
until contributions beyond the QE peak, like meson exchange currents 
and Delta effects, which may play a significant role in the analysis of data 
even at the maximum of the QE peak, are carefully 
evaluated~\cite{BCDM04,amaro05,ivanov08}. These processes contribute to the cross-section and the comparison of the 
pure nucleonic predictions of the GF1, GF2, and RMF models to data will only indicate what is the contribution
of the non-nucleonic degrees of freedom to the cross-sections.

\subsection{Scaling functions}

The effects already discussed for the differential cross sections are also 
present in the scaling functions. Here we compare results for the longitudinal 
component of the scaling function 
$f_{{L}}(\psi^{\prime})$  using the 
same descriptions for the final state interactions already considered 
for the differential cross sections. 

As a first step, the longitudinal contribution $f_{{L}}(\psi^{\prime})$  
obtained in RPWIA and rROP by the Pavia and 
the Madrid-Sevilla groups at three values of 
the momentum transfer is displayed in Fig.~\ref{fig5} showing almost coincident
results. Similar comments apply to the transverse contribution $f_{{T}}(\psi^{\prime})$.
These results, in addition to the cross sections in Fig.~\ref{fig1}, confirm the
consistency of the numerical codes when calculations are performed under the
same conditions.
  
The scaling function $f_{{L}}(\psi^{\prime})$ evaluated within RPWIA and rROP shows a very mild
dependence on the momentum transfer for both positive and negative values of the scaling variable 
$\psi^{\prime}$, i.e., violation of scaling of first kind is small.

In Figs.~\ref{fig6} and~\ref{fig7} we compare $f_L(\psi')$ evaluated with different models and
for several values of the momentum transfer with the averaged QE phenomenological 
scaling function extracted from the analysis of $(e,e^{\prime})$ data~\cite{don1,don2,mai1}. 
As already shown in previous works~\cite{cab1,cab2,isospin07}, the RMF model produces an asymmetric shape 
with a long tail in the region with $\psi^{\prime}>0$ that follows closely the phenomenological 
function behavior. The asymmetry in the data has usually been attributed to physical effects 
beyond the mean field such as two-body currents and short-range 
correlations~\cite{DePace}. Within a non-relativistic framework such contributions need to be
considered in order to get asymmetry~\cite{DePace,bleve,nieves}. By contrast,
the RMF approach is capable of explaining the asymmetric behavior of data within 
the framework of the relativistic impulse approximation taking advantage of its 
strong relativistic scalar and vector potentials. 
The results with the GF model are similar to those obtained with 
RMF at $q$ = 500 MeV$/c$ and, with moderate differences, at
$q$ = 800 MeV$/c$, while visible discrepancies appear at $q$ = 1000 MeV$/c$.
On the other hand, discussion of results for the scaling functions follows similar
trends to the one already applied to the behaviour of the cross sections in 
Fig.~\ref{fig2}, i.e., at higher $q$-values the maximum strength occurs for the GF1 model
being the RMF one the weakest.

The asymmetric shape with a tail in the region of  positive $\psi^{\prime}$ is 
obtained in both RMF and GF models which involve descriptions of 
FSI either with a strong energy-independent real potential or with a complex 
energy-dependent optical potential, respectively.
The scaling functions corresponding to RPWIA and rROP, which are also presented 
in Fig.~\ref{fig7} do not show any significant asymmetric tail for $\psi^{\prime}>0$.
The different dependence on the momentum transfer shown by the potentials 
involved in the RMF and GF approaches makes the GF scaling function tail less pronounced as the value
of $q$ goes up.

The comparison of the different models to the longitudinal scaling function is illuminating. We must recall that 
the experimental longitudinal response can be considered as a much better representation of the pure nucleonic
contribution to the inclusive cross-section than the total cross-section. It is remarkable that, as seen in figs. 6 and
7, except for the highest value of $q$ considered (1000 MeV/$c$), GF1, GF2 and RMF approaches yield very similar 
predictions for the longitudinal response, in good agreement with the experimental longitudinal response. The asymmetric tail 
of the data and the strength at the peak are fairly reproduced by the three approaches. However, for
$q=1000$ MeV/$c$, only
the RMF approach seems to be favoured from the comparison to the data, while GF1 and GF2 yield now rather different
predictions than the RMF approach, that seem to be ruled out by data. We must keep in mind that the GF approach
uses as input the phenomenological optical potentials. It is clear that, as the momentum transfer increases, the 
phenomenological optical potential will (implicitely) incorporate a larger
amount of contributions from non 
nucleonic degrees of freedom, such as for instance the loss of (elastic) flux into the inelastic delta excitation with
or without real pion production. Thus the input of the GF formalism is contaminated 
by non purely nucleonic contributions. Consequently, GF predictions
depart from the experimental QE longitudinal response, that
effectively isolates only nucleonic contributions. This difference,
which is larger with increasing momentum transfer, emerges
as an excess of strength predicted by the GF model as it translates a loss of 
flux due to non-nucleonic processes into inclusive purely nucleonic strength.
On the other hand, the RMF model uses as input the effective mean field that reproduces
saturation properties of nuclear matter and of the ground state of the nuclei involved, and thus it is much more suited to
estimate purely nucleonic contribution to the inclusive cross-section, even at $q=1000$ MeV/$c$. Taking these facts into account, the comparison of 
the models to the data, both for total cross-sections and longitudinal responses yields what one reasonably expects.   


An analysis of the scaling of first-kind, i.e., independence of the momentum 
transfer, is illustrated in Fig.~\ref{fig9}. The results are the same already
shown in Fig.~\ref{fig6}, but are presented in a different way. Each panel
corresponds to a specific description of FSI (GF1, GF2, and RMF)  
and includes results obtained for different values of $q$.
The experimental data are compatible with a mild violation of the first-kind 
scaling, particularly in the positive $\psi'$-region.
In Refs.~\cite{cab1,cab2} the scaling functions evaluated with the RPWIA and rROP 
models were shown to depend very mildly on the transferred momentum in the whole,
positive and negative, $\psi^{\prime}$ region.
In the case of the RMF approach, there is a slight shift in the region 
$\psi^{\prime}<0$, whereas the model breaks scaling approximately at $30\%$ 
level when $\psi^{\prime}>0$. Similar results are obtained with the 
GF models, where a shift in the region of negative $\psi^{\prime}$ also occurs,
and scaling is broken for $\psi^{\prime}>0$. This scaling violation for 
$\psi'>0$ is larger with GF1 due to the enhancement produced in the region of 
the QE peak by this phenomenological optical potential at higher values of $q$. 
As a consequence, the comparison between the experimental QE scaling function
and  the GF1 results is worse than the corresponding comparison for GF2.


\section{Summary and conclusions}
\label{conc}

The work developed in this paper has emerged as a close collaboration
between the Pavia and Madrid-Sevilla groups. For the last years, both
groups have been deeply involved in studies of lepton scattering
reactions with nuclei. Inclusive and exclusive electron scattering processes
have been analyzed, as well as reactions involving the use of neutrinos. The calculations
performed by the two groups, which are similar in some aspects, i.e., treatment
of relativistic ingredients, bound-nucleon wave functions, 
single-nucleon current operators, etc., present also clear
differences concerning the description of final state interactions, which 
constitute an essential ingredient for a successful comparison with data. 

The consistency of the numerical calculations developed by our two groups is checked
by comparing results in the plane wave limit (RPWIA) and
making use of the real part of the relativistic optical potential (rROP). 
Almost identical predictions come out within RPWIA and very similar ones 
with rROP. This reinforces our confidence on the reliability of both calculations. 
As known, the description of inclusive $(e,e')$ reactions requires the contribution
from the inelastic channels to be retained. Within the framework of the relativistic 
impulse approximation, a simple recipe to compute the inclusive strength is the
use of purely real potentials in the final state. This is the case of the rROP 
approach (phenomenological relativistic optical potential but without 
the imaginary part). However, although rROP conserves the flux,
its use for inclusive processes is not entirely satisfactory since the (complex) relativistic
optical potential has its origin in the description of exclusive processes where only
the elastic channel contributes to the observables. Two other models based on the 
relativistic impulse approximation have been used recently to account for FSI. The former
employes distorted waves obtained with the same relativistic mean field considered for the
initial bound states. This is denoted as RMF and it fulfills the dispersion relation and
the continuity equation. The latter procedure is based on the relativistic
Green's function (GF)
approach, which allows one to treat FSI consistently in the inclusive and exclusive
reactions.

Differential cross sections and scaling functions evaluated with both models for different
kinematical situations have been compared. Discrepancies are shown to increase with the
momentum transfer. This is linked to the energy-dependent optical potentials involved 
in the relativistic GF method by contrast to the strong energy-independent
RMF potentials. Moreover, results presented for two
different parameterizations of the ROP, namely, EDAD1 vs EDAD2, prove the 
importance of the imaginary term,
which gets its maximum strength around 500 MeV, whereas the real part gets smaller as
the energy increases. This explains the different behavior shown by  
the results of the GF and rROP models, the latter approaching the RPWIA results 
for higher values of the transferred energy and momentum. Furthermore, 
discrepancies between the GF and RMF results are also clearly visible as $q$
goes up, due to the contribution from non-purely nucleonic inelasticities to 
the phenomenological optical potential

All models considered respect scaling and superscaling properties. Furthermore,
the significant asymmetry in the scaling function produced by the RMF
is strongly supported by data~\cite{cab1}. On the contrary, asymmetry is largely absent in RPWIA
and rROP predictions. 
The relativistic GF approach leads to similar results to RMF, i.e., with 
presence of the asymmetry for intermediate $q$-values.
On the contrary, visible discrepancies emerge for larger $q$, being
the GF scaling function tail less pronounced but showing more strength in the region where
the maximum occurs. Moreover, the GF results for high $q$ present a strong dependence on the 
specific parameterization considered for the optical potential. 
These results are directly linked to effects introduced by the imaginary
(scalar and vector) term in the optical potential that presents a
high sensitivity to the particular ROP parameterization.
The relativistic GF approach, even based on the use of a complex optical potential, 
is shown to preserve flux conservation, hence the imaginary term in the potential leading
to a redistribution of the strength among different channels. This
explains the difference observed between RMF and GF predictions, the latter 
with additional strength in the region close to the maximum in the QE response.
This behavior could be
connected with effects coming from the contribution of the $\Delta$ which are,
somehow, accounted for in a phenomenological way by the GF approach, modifying 
consequently the responses even in the region where the QE peak gives the main
contribution. Notice that the higher the transfer momentum is, a stronger overlap
between the QE and $\Delta$ peaks occurs. This makes very difficult to isolate
contributions coming from either region. 

Although great caution should be exercised
in extending the above comments before more conclusive studies are performed, 
the present analysis can be helpful in disentangling different treatments of FSI
and their connections with different physics aspects
involved in the process. The similarities of the GF and RMF predictions for 
the inclusive cross-sections, particularly for intermediate values of $q$, in 
spite of the very different phenomenological ingredients they consider, and the very reasonable agreement with the data for the
longitudinal scaled response, that constitutes a good representation of the experimentally measured purely nucleonic 
response to the inclusive cross-sections, are a clear indication of the fact that both models make a
very decent job in estimating the inclusive contribution. 
It will be interesting to investigate the possibility of disentangling in the 
phenomenological optical potential the contributions due to non-nucleonic 
inelasticities and extract a 'purely nucleonic' optical potential which could
then be used in the GF approach and contrasted against the experimental 
longitudinal scaling function.
This work can be considered as a first step in this direction.


\begin{acknowledgments}

We are grateful to M.B. Barbaro for useful discussions and for her
valuable advice. This work was partially supported by DGI (Spain) under contract
nos. FPA2006-13807-C02-01, FPA2007-62216, FIS2008-04189, and the Spanish
Consolider-Ingenio 2010 programme CPAN (CSD2007-00042).
J.A.C. acknowledges support from the INFN-MEC agreement,
project ``Study of relativistic dynamics in neutrino and electron
scattering''.

\end{acknowledgments}



\end{document}